\documentclass[twocolumn]{aastex631}
\usepackage{natbib}
\usepackage{threeparttable}
\pdfoutput=1
\usepackage{subfigure}
\usepackage{amsmath}
\usepackage{epstopdf}
\usepackage{multirow}
\usepackage{pifont}
\usepackage{graphicx}

\newcommand{\refsection}[1]{Section \ref{#1}}

\begin{document}
    \title{Identifying hot subdwarf stars from photometric data using Gaussian mixture model and graph neural network}
    \author{Liu Wei}
\affiliation{ School of Mathematics and Statistics, Shandong University, Weihai, 264209,   Shandong, China}

 \author{Bu Yude}
\affiliation{ School of Mathematics and Statistics, Shandong University, Weihai, 264209,   Shandong, China}
 \email{buyude@sdu.edu.cn}

\author{Kong xiaoming}
\affiliation{ School of Mechanical, Electrical \& Information Engineering, Shandong University, Weihai, 264209,  Shandong, China}

\author{Yi Zhenping}
\affiliation{ School of Mechanical, Electrical \& Information Engineering, Shandong University, Weihai, 264209,  Shandong, China}
\author{Liu Meng}
\affiliation{ School of Mechanical, Electrical \& Information Engineering, Shandong University, Weihai, 264209,  Shandong, China}   
    
	\begin{abstract}
	    \ {Hot subdwarf stars are very important for understanding stellar evolution, stellar astrophysics, and binary star systems. Identifying more such stars can help us better understand their statistical distribution, properties, and evolution. In this paper, we present a new method to search for hot subdwarf stars in photometric data ($b$, $y$, $g$, $r$, $i$, $z$) using a machine learning algorithm,  graph neural network, and Gaussian mixture model. We use a Gaussian mixture model and Markov distance to build the graph structure, and on the graph structure, we use a graph neural network to identify hot subdwarf stars from $86\,084$ stars, when the recall, precision, and f1 score are maximized on the original, weight, and synthetic minority oversampling technique datasets. Finally, from $21\,885$
	    candidates, we selected approximately $6\,000$ stars that were the most similar to the hot subdwarf star.}
	\end{abstract}
\section{Introduction}
	Hot subdwarf stars are small-mass stars of approximately $0.47 M$ with an effective temperature of $20\,000K-80\,000K$ and logarithmic 
	surface gravities of $4.5-6.5 ~dex$. They belong to two main categories - sdO and sdB. These stars have a spectra similar to those of the main sequence $O/B$-type stars and are considered to be core helium-burning stars. In the H-R diagram, hot subdwarf stars are located at the blueward extension of the 
	horizontal branches. Thus, they  are called  extended horizontal branches \citep{1986A}.

	Hot subdwarf stars are closely related to binary systems; for example, a number of sdB stars have been discovered in binary 
	systems \citep{2010The}; hot subdwarf binaries with massive white dwarf companions are crucial to the study of hypervelocity stars \citep{Geier2015The}. Additionally, hot subdwarf stars contribute to the stellar astrophysics \citep{2012A}, test chemical diffusion theory \citep{Vick2011Abundance}, and ultraviolet upturn phenomena in elliptical galaxies \citep{2007A}. Moreover, identifying more such stars not only help us answer the above questions, but also help understand their statistical properties \citep{2011Pressure}.

	The number of known hot subdwarf stars increased when the Palomar Green survey was released \citep{1986The}. Using the Sloan Digital Sky Survey (SDSS) and Large Sky Area Multi-Object Fiber Spectroscopic Telescope Survey (LAMOST), 
	more than $5\,000$ known hot subdwarf stars were cataloged \citep{2016The}. Presently, most hot subdwarf stars have been discovered using traditional methods, such as color cuts and visual inspection. However, these methods are over-reliant on the photometric information of stars. 
  In recent years, machine learning methods have developed rapidly and methods such as hierarchical extreme learning machine algorithm have been used to identify and classify hot subdwarf stars \citep{2017Searching}. In this study, we also applied machine learning algorithms to search for hot subdwarf stars, the graph 
  neural network (GNN) and Gaussian mixture model (GMM).

	GMM is a classic unsupervised learning algorithm based on maximum likelihood parameter estimation and the expectation-maximization algorithm \citep{0Gaussian}. A typical GMM calculates the posterior probability of dividing samples 
	into clusters. It is widely used in language identification \citep{2002Approaches} and multimodal biometrics 
	verification methods \citep{2016GMM}. In this study, we apply GMM to divide hot subdwarf stars and their negative samples into six clusters and then build a graph structure.
 
	A graph is a very important data structure. For example, social networks, human skeletons, and chemical molecules are all  graph-structure data. Convolution is the most common operation in neural networks, such as convolutional neural networks (CNN), which are special deep networks that include convolution layers and pooling layers \citep{1998Gradient} and have been widely applied in semantic 
  segmentation \citep{2015Fully}, image generation \citep{2013Deep}, and video classification \citep{2016Video}. Inspired by the deep convolutional 
  network, GNN also uses convolution to extract local features on the graph, such as graph convolution networks \citep{2016Variational}, graph attention networks \citep{2018Graph}, and GraphSAGE \citep{2017Inductive}. GNN has been applied in many areas, such as 
	text classification \citep{2018Graph1}, image classification \citep{2017Few}, visual question-answering \citep{2018Out}, and intrusion detection \citep{2021Graph}. 
  Presently, GNN achieves positive results in traffic forecasting, social networks, point cloud classification, etc. In this study, we applied the GNN to classify hot  
  subdwarf stars from photometric data and present a new hot subdwarf star candidate catalog extracted from the Gaia DR2 catalog of hot subluminous stars.
    \citep{2018The}.
	 
	 The remainder of this paper is organized as follows. In \refsection{GMM}, we introduce the GMM algorithm to divide the stars into different clusters. In \refsection{GNN}, we introduce GNN algorithms including the graph convolution network, graph attention network, and GrangSAGE. After we introduced data screening and data processing in \refsection{Data Introduce}, we compared the three GNN algorithms with other main machine learning methods and obtained approximately $6,000$
	hot subdwarf star candidates with high confidence in \refsection{Data experiment}. Finally, in Section \refsection{Conclusions}, we summarize our experiments.
 
\section{GMM} \label{GMM}
	In this section, we briefly introduce the GMM. For more information on the GMM, we refer the reader to \citet{0Gaussian}. 
	
	Suppose that $X=\{x_j\}$ $(j=1,2,..,N)$ denotes the unsupervised dataset from a Gaussian distribution, where $x_j=(x_{j1}, 
	x_{j2},..., x_{jn})^T$ is the magnitude data vector with $n$ bands. 
	
	When the dataset is divided into M clusters $\Omega_1,\Omega_1,...,\Omega_M$, the GMM is given by the equation:
	\begin{equation}
		P(X|\lambda) = \sum_{i=1}^{M}\omega_ig(X|\mu_i,\Sigma_i)\
	\end{equation}
	where $\omega_i$ $(i=1,2,..,M)$ is the mixture coefficient satisfying the constraint that $\sum_{i=1}^{M}\omega_i=1$, 
	and $g(X|\mu_i,\Sigma_i)$ $(i=1,2,...,M)$ are n-variate component Gaussian densities.
	\begin{equation}
		g(X|\mu_i,\Sigma_i) = \frac{1}{(2\pi)^{n/2}|\Sigma_i|^{1/2}}e^{-\frac{1}{2}(X-\mu_i)^T\Sigma_i^{-1}(X-\mu_i)}
	\end{equation}
	with the mean vector $\mu_i$ and covariance matrix $\Sigma_i$.
	
	Suppose that the random variable $z_j\in\{1,2,..,M\}$ represents the component Gaussian densities of $x_j$ and 
	$P(z_j=i)$ represents the prior probability $x_j$ from $\Omega_i$: 
	$$P(z_j=i) = \omega_i$$
	From the Bayesian theorem, the posterior probability 
	$p(z_j=i|x_j)$ can be given by the equation
	\begin{equation}
		p(z_j=i|x_j) = \frac{\omega_iP(x_j|z_j=i)}{\sum_{l=1}^MP(z_j=l)P(x_j|z_j=l)}
	\end{equation}
	where the posterior probability $p(z_j=i|x_j)$ is parameterized as 
	$\lambda=\{\mu_i,\Sigma_i,\omega_i\} (i=1,2,...,M)$.
	
	The expectation-maximization algorithm can be applied to estimate the parameters of the GMM.
	First, initialize the parameters $\lambda$ randomly and then repeat the following two steps:
	\begin{enumerate}
		\item Calculate the posterior probability $p(z_j=i|x_j)$ to determine the cluster to which each star belongs.
		\item Based on the stars in each cluster, recalculate the parameters $\overline{\lambda}$  by maximum likelihood parameter estimation.
	\end{enumerate}
	
	When the expectation-maximization algorithm converges, the posterior probability $p(z_j=i|x_j)$ is determined, and the divided cluster $x_j$ is determined by
	\begin{equation}
		\Omega_i = \mathop{\mathrm{argmax}}\limits_{i\in\{1,2,..,M\}}p(z_j=i|x_j)
	\end{equation}
 
\section{Brief introduction of GNN}\label{GNN}
	A GNN is an artificial neural network that originates from a fully connected network based on a convolution operation. 
	A fully connected network means that the unit is connected to all units on the adjacent layer. In this study, the three GNN models used to classify hot subdwarf stars are regarded as special fully connected network models. Therefore, we first introduce a fully connected network and then introduce the GNN.
	\subsection{Fully Connected Network}
	Suppose that $\{x_i,y_i\}$ $(i=1,2,...,N)$ denotes a supervised dataset. The $x_i=(x_{i1}, x_{i2},..., x_{in})^T$ is 
	a magnitude data vector with $n$ bands. The $y_i$ is the label corresponding to $x_i$, where $y_i=1$ represents the
	hot subdwarf star, and $y_i=0$ represents other stars. 
	
  \begin{figure}[h]
    \centering
    \includegraphics[width=2.2in, height=2.0in]{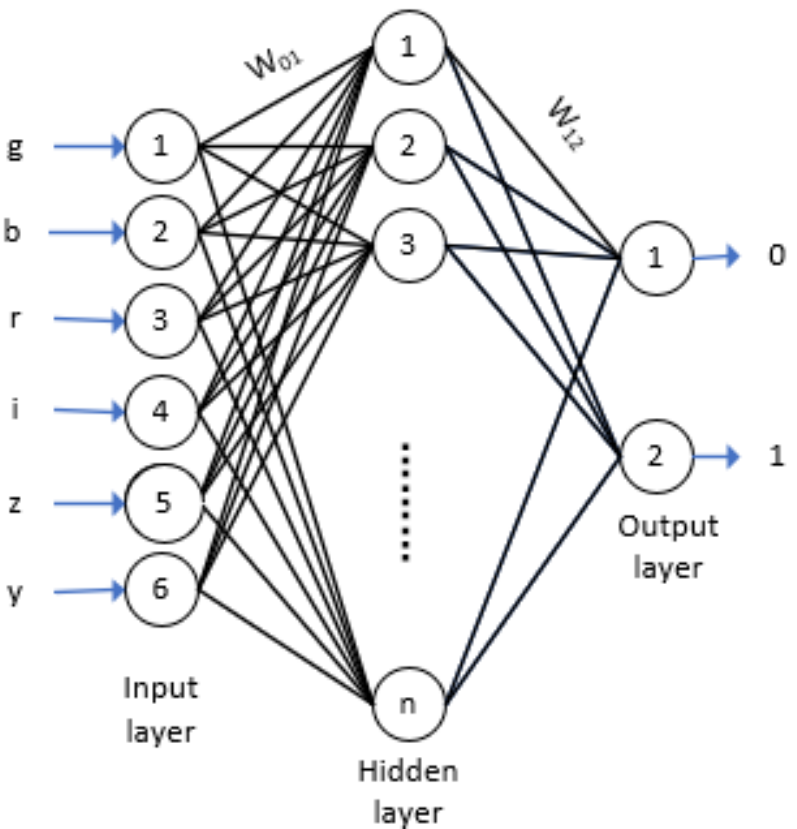}
    \caption{Architecture of the full connected network  with a one-hidden layer showing how a magnitude data vector with six bands passes through the network. In the hidden layer, the output of the $unit-j$ is 
      $\beta_j(x_i)=\sigma(\sum_{k=1}^{6}x_{ik}w_{kj}^{01})(j=1,2,...,n)$, where the $w_{kj}^{01}$ represents the weight between the 
      $unit-k$ in the input layer and the $unit-j$ in the hidden layer. In the output layer, $f(x_i)$=
      $\sigma(\sum_{j=1}^{n}\beta_jw_{j1}^{12})$, where $w_{j1}^{12}$ represents the weight between the $unit-j$ in the hidden layer and the unit in the output layer. $f(x_i)$=1 represents a hot subdwarf star, and $f(x_i)$=0 otherwise.}
    \centering
    \label{tree}
  \end{figure}
 
	For a single-layer network, the output is given by f(X)=$\sigma(W^{01}X)$. When the hidden layers are stacked, the output is expressed as 
	$L^{(j+1)}=\sigma(W^{j-1,j}L^j)$, where $L^0=X$ is the input matrix and $W^{j-1,j}$ is the weight matrix between the (j-1)th 
	and j-th layers, and $\sigma(x)$ is the nonlinear activation function. For a large dataset, a neural network with hidden layers with a nonlinear activation function can infinitely approximate any Borel measurable function. The 
	architecture is illustrated in \textbf{Figure} \ref{tree}.
 
	\begin{figure*}[htbp]
	\centering
	\includegraphics[width=6.0in, height=1.9in]{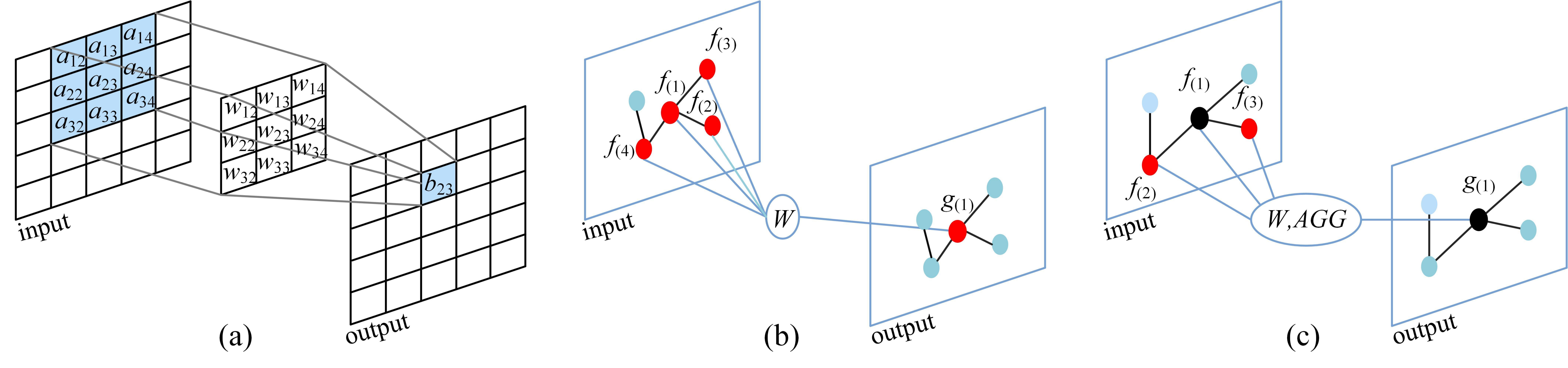}
	\caption{Panel (a): Forward propagation of CNN on the image. The $3\times3$ size convolution kernel is moved over the input layer from 
    left to right and top to bottom, so the unit in the output layer is only connected to the local unit, such that $b_{23}$ is only connected to the blue area: 
	$b_{23}=\sum_{i=1}^{3}\sum_{j=1}^{3}a_{i,j+1}w_{i,j}$. Panel (b): Forward propagation of graph convolution network and graph attention network on the graph. Both algorithms sum the features of immediate neighbor nodes by weighting, such as $g(1)=\sum_{i\in{\mathbb{N}(1)}}W_if(i)$. The difference is that the weight in the graph convolution network is the degree whereas the weight in the graph attention network is the attention coefficient. Panel (C): Forward propagation of GraghSAGE on graph. GraphSAGE randomly takes a part of the immediate neighbor nodes into aggregation, such that the node $g(1)$ in the output layer is only connected to two immediate neighbors $g_1=concat(f_1,\text{AGG}(f_2,f_3))$.}
	\centering
	\label{convolution}
	\end{figure*}	
 
	Moreover, the weighted matrix can be updated when $Loss = \sum_{i=1}^{N}(y_i-f(x_i))^2$ is minimized during the training process. We apply the error back-propagation algorithm to obtain the iteration for gradient descent, such as
	\begin{equation}
	\begin{aligned}
	\frac{\partial{Loss}}{\partial{w_{j1}^{12}}}&=\frac{\partial{Loss}}{\partial{f(x_i)}}\frac{\partial{f(x_i)}}
	{\partial{b_i}}\frac{\partial{b_i}}{\partial{w_{j1}^{12}}}\\
	&=f(x_i)(1-f(x_i))(y_i-f(x_i))\beta_j(x_i)
	\end{aligned}
	\end{equation}
	where $b_i=\sum_{j=1}^{n}\beta_jw_{j1}^{12}$ denotes the input to the output layer. 
	
	For more information about fully connected networks and the error backpropagation algorithm, we refer the reader to \citet{1986Learning}.
	
	\subsection{GNN}
	In the graph structure, all GNNs first aggregate the adjacent node information by convolution, and then input the aggregated information into the fully connected network model. Therefore, we first introduce the graph and convolution followed by three GNN models:  graph neural network, graph attention network, and GraphSAGE.
	
	\textbf{Graph structure.} A graph is an ordered triple $(V(G),E(G),A)$ with a non-empty set of vertices $V(G)=(v_1,v_2,...,v_N)$, a set of edges
	$E(G)$$\subseteq$$V$$\times$$V$, and the symmetric adjacency matrix $A$ = [$a_{ij}$], where $a_{ij}=1$ if there exists an 
	edge from $v_i$ to $v_j$ or $a_{ij}=0$ if there is no edge from $v_i$ to $v_j$. Degree matrix of adjacency 
	matrix is $D = Diag(D_{ii})$, where $D_{ii}=\sum_{j=1}^nA_{ij}$. If two nodes are directly connected to the graph, they are called immediate neighbors. For node classification, stars correspond to the nodes and the magnitudes of the star correspond to node features.
	
	\textbf{Convolution operation.}
	The convolution operation refers to the weighted summation of a sequence, such as
	\begin{equation}
		(f*g)(x) = \sum_{\tau=-\infty}^{\infty}f(\tau)g(x-\tau)
	\end{equation}
	where $g$ is a function sequence and $f$ is the weight sequence. 
 
	Convolution can efficiently aggregate the features of adjacent nodes in images and graphs. Aggregating feature 
	refers to the weighted sum of the adjacent features in the next layer. Sparse interactions refer to a field smaller than the input that interacts with the next layer. With sparse interactions, the number of model parameters can be reduced and the network depth can be increased, thereby improving the generalization level. The most popular convolution operation is the CNN. Panel (a) of \textbf{Figure} \ref{convolution} illustrates the operation of the CNN method on an image.
  
  \textbf{Graph convolution network.} For a one-layer graph convolution network, the strategy to aggregate neighbor features is to sum the features of all immediate neighbors with a weight equal to the reciprocal of the degrees. The aggregated features of node v can be expressed as
  \begin{equation}
  	x^k_v = \sigma(\sum_{j\in \mathbb{N}(v)}\frac{x^{k-1}_j}{D_j}+x^{k-1}_v)
  \end{equation}
  where $x^k_v\in \mathbb{R}^m$ is the feature of node v after k aggregations, $\mathbb{N}(v)$ is the set of immediate neighbors of node v,  $D_j$ is the degree of node j, and $\sigma$ is the activation function.
  
  Panel (b) of \textbf{Figure} \ref{convolution} illustrates the operation of the GCN on a graph.
  
  \textbf{Graph attention network.} The strategy of the graph attention network that aggregates neighbor features is to calculate the attention coefficient, and then use the attention coefficient as a weight to sum the immediate neighbors. The attention coefficient can measure the level of correlation between the nodes and is given as
  \begin{equation}
  e_{ij}=a(\text{concat}(Wx^{k-1}_v,Wx^{k-1}_v))
  \end{equation}
  where the attention coefficient $e_{ij}$ indicates the importance of node j to node i, $a$ is usually a one-layer fully neural network, $W \in \mathbb{R}^{n\times{m}}$ converts the input features into higher-level features because $n>m$.
 
  To make the attention weights easily comparable between different nodes, the graph attention network uses the softmax function to normalize all $e_{ij}$:
  \begin{equation}
  \alpha_{ij} = \frac{exp(e_{ij})}{\sum_{k\in \mathbb{N}(i)}exp(e_{ik})}\ \ \ \ \ for \  \forall j\in{\mathbb{N}(i)}
  \end{equation}
  
  Finally, the aggregated features of node i can be expressed as
  \begin{equation}
  	x^k_i = \sigma(\sum_{j\in \mathbb{N}(i)}\alpha_{ij}Wx^{k-1}_i)
  \end{equation}
  
   Panel (b) of \textbf{Figure} \ref{convolution} also illustrates the operation of the graph attention network.
  
  \textbf{GraphSAGE.}  Introduced by the graph convolution network and graph attention network, GraphSAGE also aggregates the information of immediate neighborhoods. GraphSAGE forward propagation is given as
  \begin{equation}
  \begin{aligned}
    &x^k_{\mathbb{N}(v)} = \textbf{agg}_k(\{x^{k-1}_u, \forall u \in \mathbb{N}(v)\})\\
    &x^k_v = \sigma(W \cdot \text{concat}(x^{k-1}_v,x^k_{\mathbb{N}(v)}))
  \end{aligned}
  \end{equation}
  where \textbf{agg}$_k$ represents the aggregation strategy in the $k$-layer. The common strategies in GraphSAGE are pool, mean, and lstm. Owing to the concat operation, $x^k$ has 2m dimensions. Unlike the graph convolution network and graph attention network, GraphsSAGE abandons the aggregation of all immediate neighbor features but selects a fixed number of neighbors randomly for aggregation. Panel (c) of \textbf{Figure} \ref{convolution} illustrates the operation of GraghSAGE.
  
  All three GNNs aggregate immediately neighbor information by convolution, indicating that the GNN models are sparsely connected. When multiple GNN layers are stacked, the features of the further nodes are integrated, which is called receptive-field expansion.
\section{Data Introduction} \label{Data Introduce}
  In this section, we introduce the data sources and process data.  
  \\ \indent Our dataset consists of $g, b, r, i, z, y$ band photometry of four classes of stars: hot subdwarf stars, hot subdwarf star candidates, blue horizontal-branch stars (BHB), and unknown types of stars. 
  The $g,b,r$ bands are taken from Gaia DR2. \citet{2017Gaia} introduced Gaia DR2 in more detail. Gaia was launched from Kourou on a Soyouz-Fregat rocket;  the Gaia survey used two telescopes and both feed three instruments (astrometric instrument, spectrophotometer, and spectrograph). To reduce systematics and improve precision, Gaia DR2 has optimized data processing  pipelines and calibration of instruments, and it obtains white-light G-band (330--1050 nm) magnitudes of approximately 1.7 billion sources, and blue (BP, 330--680 nm) and red (RP, 630-1050 nm) band magnitudes of approximately 1.3 billion. The photometry data $i, z, y$ bands were taken from PanStarrs DR1; \citet{chambers2016pan} introduced PanStarrs DR1 in more detail. PanStarrs has the largest digital cameras in the world, each with almost 1.5 billion pixels, and covers the entire sky north of a -30 $^\circ$ declination at least 60 times. The optical design of PanStarrs includes a wide field Ritchey--Chretien configuration with a 1.8 m primary mirror and 0.9 m secondary mirror; the wavelength range of $i, z, y$ band is 677.8--830.4 nm, 802.8--934.6 nm, 911.0--1083.9 nm respectively.
 
  Hot subdwarf stars are positive samples from \citet{2016The}; the hot subdwarf star candidates are unlabeled samples to be searched from \citet{2018The}. Negative samples are BHB stars and unknown stars, where BHB stars are from \citet{2021A} and unknown stars are randomly selected following the criterion:
  
  \begin{scriptsize}
  $\left.
  \begin{array}{ll}
    ({\romannumeral 1}) & -0.5 \ge \text{bp-rp} \ge 0.4\\
    ({\romannumeral 2}) & 2 \ge \text{M}_\text{G} \ge 6\\
    ({\romannumeral 3}) & \text{M}_\text{G} \ge \text{10(bp-rp)}+2
  \end{array}
  \right\}(\text{Selection criterion})$
  
  \end{scriptsize} 
where $\text{M}_\text{G}$ is the absolute magnitude, $\text{M}_\text{G}=\text{G}+5+5log(parallax/1000)$. The selection criterion covers the area where hot subdwarf stars are likely to appear on the Gaia DR2 Hertzsprung--Russell diagram, which can make negative samples more similar to hot subdwarf stars and improve the persuasion of our model \citep{2018Gaia,2018New}. 
	
	\begin{table}[htbp]
	\caption{Distribution of dataset}{}	
	\centering\hspace{-26pt}
	\begin{tabular}{ccccc}
		\hline\hline
		\multicolumn{2}{l}{~~~~~~~Samples}&&\multicolumn{2}{l}{~~~~~~~~~~~~~~~~~~~Stars}\\
		\cline{1-2}
		\cline{4-5}
		\multirow{2}{*}{negative}&\multirow{2}*{$59\,884$}&&unknown&$46\,509$\\
		&&&BHB&$13\,375$\\
		\cline{1-2}
		\cline{4-5}
		positive&$4\,315$&&hot subdwarf star&$4\,315$\\		
		\hline
		\multicolumn{5}{l}{\bf Notes.} \\
		\multicolumn{5}{l}{The negative samples include the unknown stars and }\\
		\multicolumn{5}{l}{BHB. The positive samples are the hot subdwarf stars.}\\
	\end{tabular}
	\label{tt1}
\end{table}.

	After the above selection, the dataset contains $64\,199$ stars, as shown in \textbf{Table} \ref{tt1} and $21\,885$ candidates. It is obvious that our dataset is unbalanced, and there are far more negative samples than positive samples. Therefore, hot subdwarf stars selected from the candidates were more reliable. 	
	
	In each experiment, we used the following three steps to preprocess the entire  dataset.
  \begin{enumerate}
    \item Normalizing the photometry data using the formula:
          $$x=x\bigg/\sqrt{\sum_{i=1}^{N}x_i^2}$$
     where $x=(x_1,x_2,...,x_N)$ is a band.
  	\item Using the synthetic minority oversampling technique (SMOTE) and increasing the weight of hot subdwarf stars to 
  	resolve the problem of unbalanced datasets. The SMOTE is an oversampling method that generates new hot subdwarf star samples
  	between two hotsubdwarf stars. Increasing the weight means that the model is more inclined to predict the candidates as hot subdwarf stars. We denote the original data as original, the data obtained by the increasing weight method as weight, and the data obtained by the SMOTE method as SMOTE. The size of original and weight dataset is equal and less than the SMOTE dataset.
  	\item Randomly divide the positive and negative data into two parts in the ratio 7:3, one for training and the other for testing.
	\end{enumerate}

\section{ Experiment}\label{Data experiment}	
	In this section, we first construct graphs from the color magnitude diagram (CMD), use the graphs to train our model and search for
	hot subdwarf stars from the candidates. Subsequently, we compared different datasets and activation functions to obtain the best 
	GNN algorithm. In addition, we compare the graph algorithm with another mainstream non-GNN algorithm. 
	
 \subsection{Model optimization and performance measure}
	We designed a three-layer neural network model with the first and second layers as the GNN model (graph convolution network, graph attention network, and GraphSAGE); the third layer was a fully connected layer. In the input layer, the six units correspond to six bands and two hidden layers with 20 and 15 units, respectively, to make a 2-layer GNN.
	The photometric message passes among the nodes in two steps, and a fully connected layer is used to increase the depth. In our experiment, we found that more layers and units did not improve the performance or slow down the training.  As it is the GNN that plays the main role, we still refer to the model as GNN, and use the cross-entropy 
	error as a loss function.
	$$L=-\sum_{i=1}^{N}y_ilogy_i+(1-\hat{y}_i)log(1-\hat{y}_i)$$.
	where $y_i$ is the true value, $\hat{y}_i$ is the predicted value that can take values of only zero or one (corresponding to hot
	subdwarf stars).
	
	\textbf{Model performance.} The most common performance measure is accuracy, which is the proportion of all correct predictions. The number 
	of hot subdwarf stars is small; therefore, when the model predicts that not all samples are hot subdwarf stars, it will still have 
	high accuracy; however, it qualifies as a poor model. Therefore, the values of recall, precision, and f1 can be used to quantify the results of our 
	model. We set
\begin{itemize}
	\setlength{\itemsep}{0pt}
	\setlength{\parsep}{0pt}
	\setlength{\parskip}{0pt}
	\item $TP$ is the number of all hot subdwarf stars correctly classified by the model as hot subdwarf stars.
	\item $FN$ is the number of all hot subdwarf stars incorrectly classified by the model as not hot subdwarf stars.
	\item $FP$ is the number of other stars incorrectly classified by the model as hot subdwarf stars.
\end{itemize}
	Then, the recall score is defined as
	$$recall = \frac{TP}{TP+FP}$$,
	The precision score is defined as
	$$precision = \frac{TP}{TP+FN}$$.
	The f1 score is defined as
	$$f1 = \frac{2precision*recall}{precision+recall}$$.

\begin{figure*}[htbp]
	\centering
	\includegraphics[width=5in, height=3.2in]{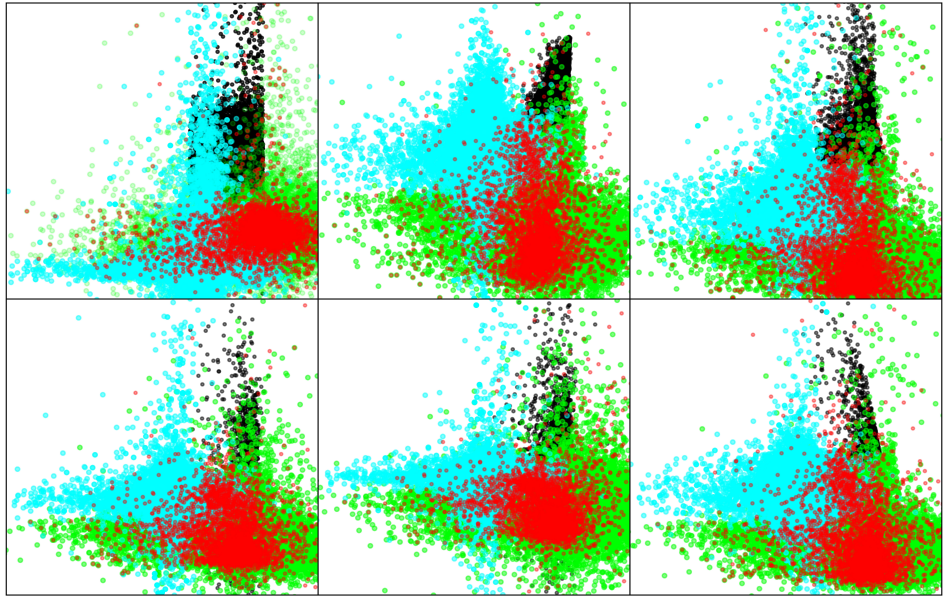}
	\caption{Color-magnitude diagram. The horizontal and vertical axes are the absolute magnitude and color index, respectively. The red dots represent hot subdwarf stars, the green dots represent hot subdwarf star candidates, the black dots represent unknown stars, and the sky blue dots represent BHB.}
	\centering
	\label{cmd}
\end{figure*}

	\textbf{Construct graph}. We built two large and homogeneous graphs whose nodes contained four stars in our dataset: one named $G_1$ for training the original and weight dataset, and the other named $G_2$ for training the SMOTE dataset. The number of nodes of $G_1$ and $G_2$ are the sizes of the original (or weight) and SMOTE datasets, respectively. We built the edges among stars in the two graphs based on the GMM and Markov distance in the following three steps:
	\begin{enumerate}
		\item Dividing the hot subdwarf stars, BHB, and unknown stars into six clusters. Two stars in the same cluster implies a high statistical similarity, whereas those in different clusters indicate little correlation. Therefore, we do not add edges between star pairs in different clusters. The result of GMM is shown in \textbf{Table} \ref{t1}. We find that most hot subdwarf stars are divided into the same cluster, which helps to reduce the number of edges between positive and negative samples.
		\item Calculating the Markov distance on six CMD, the CMD is shown in \textbf{Figure} \ref{cmd}. On the CMD, the same color dots are concentrated together, which is advantageous in constructing a graph. In $C_1$, we add an edge between two stars when the distance is less than 0.15, and in other clusters, we add an edge between the two stars when the distance is less than 0.35, which enables the adjacency matrix to be sparse and improves the operation speed without reducing the accuracy.
		\item Calculating the Markov distance between the hot subdwarf star candidates and other stars, we add an edge 
		when the distance is less than 0.25, which ensures that the candidates are connected to similar stars.
	\end{enumerate}
	
		\begin{table}[htbp]
		\caption{Result of GMM on three datasets}{}	
		\centering
		\begin{tabular}{cccccc}
			\hline\hline
			\multirow{2}{*}{Cluster}& \multicolumn{2}{c}{$G_1$}&& \multicolumn{2}{c}{$G_2$}\\
			\cline{2-3}
			\cline{5-6}
			& negative&positive& & negative & positive \\
			\hline
			$C_1$&244&2934&&31&7557\\	 	
			$C_2$&11664&233&&15026&52\\	 
			$C_3$&31689&11&&10408&1150\\
			$C_4$&747&38&&2683&247\\
			$C_5$&11111&579&&20101&0\\
			$C_6$&4677&272&&11883&3020\\
			\hline
			\multicolumn{6}{l}{\bf Notes.} \\
			\multicolumn{6}{l}{The $C_i$ represents the i-th cluster.}\\
		\end{tabular}
		\label{t1}
	\end{table}.
	
\begin{figure*}[htbp]
	\centering
	\includegraphics[width=5.0in, height=2.0in]{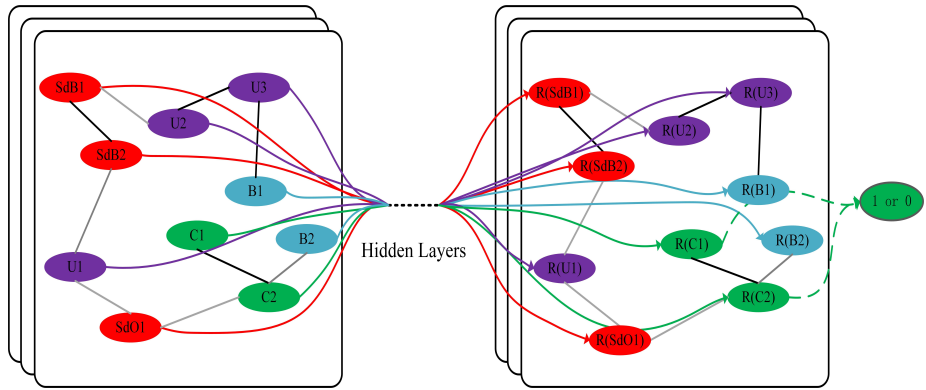}
	\caption{Schematic of stellar graph. The red nodes are hot subdwarf stars labeled 1, the purple and blue nodes are the 
		unknown stars and BHB labeled 0, and the green nodes are hot subdwarf star candidates without labels. The black edges connect the nodes with the same labels, and the gray edge connects the nodes with different labels. The $R(x)$ is the graph-aggregated features of $x$. Finally, the GNN predicts the labels of hot subdwarf stars by $R(x)$.
	}
	\centering
	\label{graph2}
\end{figure*}

	We call the graph whose nodes are stars, and the edges are built through the above three-step stellar graph. A schematic of the stellar graph is presented in \textbf{Figure} \ref{graph2}.

 \subsection{Training GNN}
 	We first introduce the experiments of training the graph convolution network, graph attention network, and GraphSAGE on three datasets. All the trained algorithms are based on Adam with a maximum of 1500 epochs; training is stopped if the cross-entropy error function did not decrease for five consecutive epochs. One of the most important parameters in a GNN algorithm is the activation function, $\sigma(zw_j)$, where $z$ is the aggregated feature on the graph. The activation functions can be classified as follows:
 
 \begin{table*} 
   \caption{Performance comparison of GNN algorithms with different activation functions}{}
   \centering
   \begin{tabular}{cccccccccc}
     \hline\hline
     \multicolumn{2}{c}{\multirow{2}*{Algorithm}}& \multicolumn{2}{c}{original} & & \multicolumn{2}{c}{weight}& & \multicolumn{2}{c}{SMOTE}\\
     \cline{3-4}
     \cline{6-7}
     \cline{9-10}
     \multicolumn{2}{c}{~}& ~~~precision~~~ & ~~~recall~~~& & ~~~precision~~~ & ~~~recall~~& & ~~~precision~~~ &~~~recall~~~\\
     \hline
     ~sigmiod   &~~~GAT~~~& ~~~0.914~~~& ~~~0.898~~~&& ~~~0.829~~~& ~~~0.940~~~ && ~~~0.968~~~ &~~~0.950~~~\\
     &  GCN     & ~~~0.910~~~ & ~~~0.810~~~ && ~~~0.791~~~ & ~~~0.880~~~ && ~~~0.973~~~ &~~~0.944~~~\\
     &   GraphSAGE & ~~~\textbf{0.930}~~~ & ~~~\textbf{0.897}~~~ && ~~~\textbf{0.831}~~~ & ~~~\textbf{0.953}~~~ && ~~~0.960~~~ &~~~0.955~~~\\
     \hline
     ~tanh &~~~GAT~~~& ~~~0.899~~~ &~~~0.815~~~ && ~~~0.866~~~ & ~~~0.941~~~& & ~~~0.964~~~ &~~~0.963~~~\\
     &   GCN  & ~~0.900~~ & ~~0.802~~& & ~~0.739~~ & ~~0.885~~ && ~~0.976~~ &~~0.847~~\\
     &   GraphSAGE  & ~~0.918~~ & ~~0.903~~& & ~~0.840~~ & ~~0.935~~& & ~~0.964~~ &~~0.957~~\\
     \hline
     ~relu~~&~~GAT~~~& ~~0.917~~ &~~0.907~~&&~~0.637~~&~~0.938~~&&~~\textbf{0.966}~~&~~\textbf{0.961}~~\\
     &   GCN   & ~~0.883~~ & ~~0.817~~ && ~~0.727~~ & ~~0.888~~& & ~~0.947~~ &~~0.949~~\\
     &   GraphSAGE  & ~~0.923~~ & ~~0.890~~ && ~~0.881~~ & ~~0.927~~ && ~~0.960~~ &~~0.943~~\\
     \hline
     ~SoftPlus   &  GAT  & ~~0.921~~ & ~~0.877~~ && ~~0.729~~ & ~~0.937~~ && ~~0.966~~ &~~0.949~~\\
     &   GCN   & ~~0.910~~ & ~~0.806~~& & ~~0.822~~ & ~~0.865~~ && ~~0.953~~ &~~0.952~~\\
     &   GraphSAGE  & ~~0.910~~ & ~~0.895~~ && ~~0.821~~ & ~~0.938~~& & ~~0.958~~ &~~0.954~~\\
     \hline
     ~sgn  &    GAT  & ~~0.833~~ & ~~0.552~~ && ~~\ding{53}~~ & ~~0.803~~ && ~~0.804~~ &~~0.649~~\\
     &   GCN   & ~~0.751~~ & ~~0.696~~& & ~~0.669~~ & ~~0.811~~  && ~~0.687~~ &~~0.727~~\\
     &   GraphSAGE   & ~~\ding{53}~~ & ~~\ding{53}~~ && ~~0.358~~ & ~~0.861~~ && ~~0.813~~ &~~0.652~~\\
     \hline
     \multicolumn{10}{l}{\bf Notes.} \\
     \multicolumn{10}{l}{Results of three GNN algorithms with different activation functions on three data sets. These experiments were}\\ 
     \multicolumn{10}{l}{conducted to assess the ability of GNN algorithms and activation functions to classify hot subdwarf stars. GAT}\\
     \multicolumn{10}{l}{and GCN represent the graph attention network graph convolution network, respectively. Bold represents the b-}\\
     \multicolumn{10}{l}{est performance, and \ding{53} means that model gets 0 score.}\\         
   \end{tabular}
   \centering
   \label{tab}
  \end{table*}

 	\begin{enumerate}
 	\item $sigmiod$	function
 	$$\sigma(zw_j)=1/(1+e^{zw_j})$$
 	\item $tanh$ function
 	$$\sigma(zw_j)=(e^{zw_j}-e^{-zw_j})/(e^{zw_j}+e^{-zw_j})$$
 	\item $relu$ function
 	$$\sigma(zw_j)=max(0,zw_j)$$
 	\item $SoftPlus$ function
  $$\sigma(zw_j)=ln(1+e^{zw_j})$$
  \item $sgn$ function
	$$\sigma(zw_j)=\left\{
   \begin{aligned}
  	&1,zw_j>0\\
  	&-1,otherwise.
 \end{aligned}
   \right.$$ 
   \end{enumerate}

  The results of using the original, weighted, and SMOTE datasets are shown in \textbf{Table} \ref{tab}. We used different types of activation functions to determine the type of function that performed best. From the table, we find that the original dataset is more likely to give a high precision score, the weighted dataset is more likely to give a high recall score, and the SMOTE dataset provides a high recall and precision score. Because there is no case where the recall and precision scores reach the highest at the same time, we take f1 as the measure of performance.  Furthermore, the GraphSAGE with the $sigmoid$ function gives the best performance on the original and weighted datasets, and the graph attention network with the $relu$ function gives the best results on the SMOTE dataset.
  
  Another important parameter of the GraphSAGE algorithm is the aggregator. Aggregator introduces the aggregation of neighbor features, and in some datasets, diverse aggregators have distinct classification capabilities. The aggregator can be classified as follows:

 \begin{enumerate}
  \item $mean$ aggregator. It concatenates $v$ and the mean of vectors in \{$h_v^{k-1},\forall{u}\in{N(v)}$\}. 
  $$h_v^k = \sigma(\textbf{W}\cdot\text{MEAN}(\{h_v^{k-1}\}\cup\{h_u^{k-1},\forall{u}\in{N(v)}\}))$$
  \item $lstm$ aggregator. LSTM is a recurrent neural network and applying the LSTM to random permutation of the node’s neighbors can aggregate the feature of
   the vectors in \{$h_v^{k-1},\forall{u}\in{N(v)}$\}; \citet{6795963} introduces LSTM in more detail.
  \item $pooling$ aggregator. It concatenates $v$ and maximizes pooling of vectors in \{$h_v^{k-1},\forall{u}\in{N(v)}$\}:
    $$\text{AGG}_K^{\text{pool}} = max(\{\sigma(\textbf{W}_\textbf{pool}h_u^k+b),\forall{u}\in{N(v)}\})$$
 \end{enumerate}

\begin{table*}[htbp]
	\caption{Performance comparison of different aggregators}{}
	\centering
	\begin{tabular}{ccccccccccccc}
		\hline\hline
		\multicolumn{2}{c}{\multirow{2}*{Algorithm}}& \multicolumn{3}{c}{original}&&\multicolumn{3}{c}{weight}&&\multicolumn{3}{c}{SMOTE}\\
		\cline{3-5}
		\cline{7-9}
		\cline{11-13}
		&& precision& ~recall~&~f1~ && ~precision~ & ~recall~&~f1~ && ~precision~ & ~recall~&~f1~\\
		\hline
		sigmiod&~mean & ~\textbf{0.930}~ & ~\textbf{0.897}~&~\textbf{0.913}~ && ~0.831~ & ~0.953~&0.886~ && ~0.960~ & ~0.955~&0.958\\
		&~lstm & ~0.901~ & ~0.893~&0.897 && ~0.806~ & ~0.941~&0.868 & & ~0.972~ & ~0.963~&0.968\\  
		&~pool & ~\ding{53}~ & ~\ding{53}~&\ding{53}~ && ~0.778~ & ~0.944~&~0.853~ && ~0.971~ & ~0.960~&0.966\\
		\hline
		relu&~mean & ~0.923~ & ~0.900~&0.911~ && ~\textbf{0.881}~ & ~\textbf{0.927}~&~\textbf{0.940}~ && ~0.961~ & ~0.943~&0.952\\
		&~lstm & ~0.894~ & ~0.894~&0.894~&& ~0.860~ & ~0.882~&0.871~ & & ~\textbf{0.975}~ & ~\textbf{0.966}~&\textbf{0.972}\\  
		&~pool & ~0.837~ & ~0.928~&0.900~ && ~0.785~ & ~0.952~&0.860~ && ~0.959~ & ~0.960~&0.955\\
		\hline    
		\multicolumn{13}{l}{\bf Notes.}\\
		\multicolumn{13}{l}{Results of GraphSAGE with sigmoid and relu function on three data sets. These experiments were conducted to ass-}\\ 
		\multicolumn{13}{l}{ess the ability of GraphSAGE algorithms with different aggregator to classify hot subdwarf stars. Bold represents the}\\
		\multicolumn{13}{l}{best performance, and \ding{53} means that GraphSAGE gets 0 score.}\\
	\end{tabular}
	\centering
	\label{tab11}
\end{table*}

\begin{figure*}
    \centering
    \includegraphics[width=5.4in, height=2.1in]{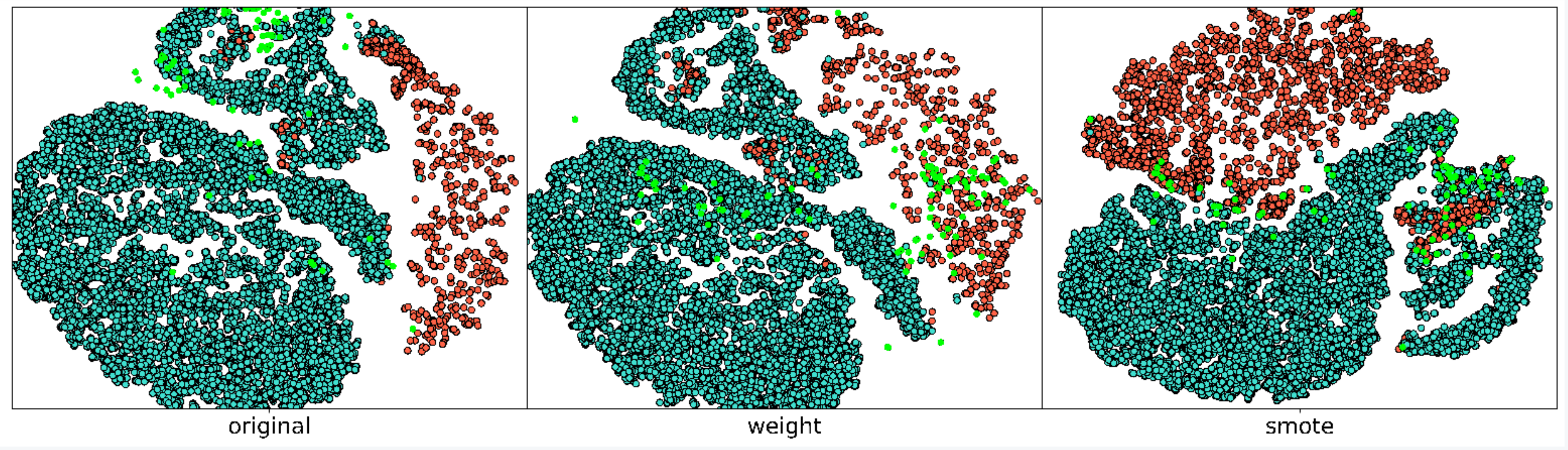}
    \caption{ t-sne visualization of test dataset. The blue dots are the negative samples, the red dots are the
    	positive samples, and the green dots indicate the samples wrongly classified by GraphSAGE. 
    }
    \label{sgsg}
	\end{figure*}

  \textbf{Table} \ref{tab11} shows the results of GraphSAGE with $sigmoid$ and $relu$ function. From the table, we find that, on the original dataset, the mean aggregator gives the best results, and on the weight and SMOTE datasets, the $lstm$ aggregator gives the best result. Comparing \textbf{Table} \ref{tab} and \textbf{Table} \ref{tab11}, we found that the GraphSAGE algorithm gives better performance than the graph attention network and graph convolution network. Hence, in our experiments, we used the GraphSAGE algorithm as the best-performing algorithm to classify hot subdwarf stars. We also provided a visualization result of GraphSAGE. We use the t-sne tool \citep{Laurens2012Visualizing} to transform the $b,y,g,r,i,z$ bands into two dimensions and then visualize the distribution of the prediction on the test set. \textbf{Figure} \ref{sgsg} shows that positive and negative samples are clearly distinguished by GraphSAGE and the number of samples wrongly classified by GraphSAGE are small. Furthermore, because of the increased weight and sample generation, there are more red dots on the weight and SMOTE datasets than in the original. In the SMOTE dataset, there are few incorrectly classified samples; most of them appear on the boundary, which indicates that the SMOTE dataset is better for GraphSAGE.
 
\subsection{ Comparison of GraphSAGE with other popular algorithms}

	\begin{table*}[t]
		\centering
		\caption{Model configurations using Bayesian optimization}{}
		\begin{tabular}{ccccccc}
			\hline\hline
			& hyperparameter & set & best & hyperparameter & set & best\\
			\hline
			lightgbm&n\_estimators&(100,400)&[379,467,379]&max\_depth&(1,8)&[6,6,4]\\
			&min\_child\_samples&(2,6)&[6,4,6]&&&\\
			\hline
			svm&C&(0.1,$10^3$)&[878,875,868]&kernel&[rbf,sigmiod, &[rbf\\
			 &gamma&(0,2)&[2,1.9,1]&& poly,linear]& rbf,rbf] \\
			\hline
			logistic& C&(1,10)&[5,4,9]&penalty&(l1,l2)&[l2,l2,l1]\\
			\hline
			Ridge& alpha&(0.5,1) &[0.7,0.6,0.2]&&&\\
			\hline
			passive-aggressive& C& (1,10)& [7,9,0.4] &&&\\
			\hline       
			\multicolumn{7}{l}{\bf Notes.} \\
			\multicolumn{7}{l}{Set represents the range of the hyperparameters, Best indicates the best value of the model for original weight, and}\\ 
			\multicolumn{7}{l}{SMOTE dataset.}\\
		\end{tabular}
		\label{tab1}
	\end{table*}  
 
To further assess the performance of GraphSAGE, we compared GraphSAGE with five popular algorithms: passive-aggressive algorithm, support vector machine (svm), 
logistic regression, ridge classifier, and light gradient boosting machine (lightGBM). These five algorithms have wide 
applications in classification missions.

The logistic regression, passive-aggressive, and ridge classifier are known as weak learners because they all have only one layer. Logistic regression applies a sigmoid function to a linear model and classifies the unlabeled sample by providing the probability of the samples belonging to each class. The ridge classifier adds an $L_1$-norm parameter to the loss function of the linear model to prevent overfitting. The passive-aggressive algorithm is an online-learning algorithm in which the passive operation refers to changing the loss function if the sample classification is wrong, and aggressive function refers to replacing the learning rate with the loss function.

The svm divides the dataset by a hyperplane and introduces the kernel method to enhance the weak learner. The most common kernels include $sigmoid, RBF, poly,$ and $linear$. For classifying hot subdwarf stars, the $RBF$ kernel yields the best performance and is given as
$$K(x_1,x_2)=exp(-\|x_1-x_2\|^2/\sigma)$$ where $\sigma$ is the scale parameter to be optimized.

The lightGBM is an evolutionary process of GDBT. The GDBT is an ensemble algorithm based on stacking multiple decision trees as weaker learners. The weaker learner will give a higher weight to the samples wrongly divided by the previous weaker learner. The final result of GBDT is the weighted sum of each weaker learner. Based on the GBDT, lightGBM samples the training data, abandoning many samples with a small gradient. Therefore, lightGBM is faster than GBDT; however, its performance is not reduced.

The above five algorithms are widely used in different fields; therefore, we compare the performance of these algorithms with GraphSAGE. To improve the robustness, we search for the best hyperparameters using Bayesian optimization; the results are shown in \textbf{Table} \ref{tab1} and \textbf{Table} \ref{t5}, respectively. From the table, we found that GraphSAGE+SMOTE gave the highest scores compared to the other combinations. Hence, GraphSAGE+SMOTE is a better alternative than other machine-learning algorithms and datasets for searching for hot subdwarf stars.

 We applied the GraphSAGE algorithm to the $b, y, g, r, i, z$ bands of the SMOTE dataset to classify hot subdwarf star candidates. To increase the reliability, we added a threshold of 12 to the output layer when the hot 
subdwarf star candidates were inputted into GraphSAGE. Moreover, GraphSAGE is more inclined to predict candidates as negative samples, and the hot subdwarf stars selected by GraphSAGE were more confident. Finally, we found approximately $6\,119$ hot subdwarf star candidates with a higher confidence.
 
	\begin{table}[htbp]
	\centering
	\caption{ f1 score of different models}{}
		\begin{tabular}{c|ccc}
			\hline\hline
			&original&weight&SMOTE\\
			\cline{1-4}
			GraphSAGE&0.913&0.940&\textbf{0.972}\\
			lightGBM&0.922&0.928&0.961\\
			svm&0.900&0.897&0.863\\
			logistic&0.634&0.673&0.800\\
			ridge&\ding{53}&0.512&0.339\\
			passive-aggressive&0.681&0.688&0.734\\
			\cline{1-4}
			\multicolumn{4}{l}{\bf Notes.}\\
			\multicolumn{4}{l}{Results of various algorithms on three data sets.}\\
			\multicolumn{4}{l}{\ding{53} means that ridge gives 0 score, and bold repr-}\\
			\multicolumn{4}{l}{esents the best performance.}\\
		\end{tabular}
	\centering
	\label{t5}
	\end{table}
 
\section{Conclusions}\label{Conclusions}
	In this paper, we propose a novel deep learning algorithm to find hot subdwarf stars based on $b$, $y$, $g$, $r$, $i$, and $z$ bands and the GraphSAGE algorithm and use the SMOTE method to process unbalanced data.
	
	SMOTE is an oversampling method used to resolve unbalanced datasets. As the search for hot subdwarf stars involves working with unbalanced data, SMOTE was applied in the experiments.
	
	GraphSAGE is a hierarchical learning framework for a graph neural network and graph attention network algorithm. In contrast to traditional machine learning, GraphSAGE classifies hot subdwarf stars on a graph, which is built from a color-magnitude diagram using the Markov distance, obtaining the GMM edges between nodes. There are two important hyperparameters in GraphSAGE. One is the activation function and the other is the aggregator. In our experiment, $relu + lstm$ showed the best performance.
	
	We compared GraphSAGE with two GNN models and five machine learning algorithms, and the results showed that GraphSAGE provides
	the best results. Finally, we predict $21\,885$ hot subdwarf star candidates and select $6\,119$ stars that are most similar to hot subdwarf stars.

     This work is supported by the National Natural  Science Foundation of China (NSFC) under Grant Nos. 11873037, U1931209, and 11803016, the science research grants from the China Manned Space Project with No. CMS-CSST-2021-B05 and CMS-CSST-2021-A08, 
    and is partially supported by the Young Scholars Program of Shandong University, Weihai (2016WHWLJH09).

\bibliographystyle{aasjournal}
\bibliography{refs}

\end{document}